\documentclass[prl,twocolumn,superscriptaddress,showpacs,preprintnumbers] {revtex4}

\usepackage{amssymb,amsmath,amsfonts}
\usepackage{graphicx}

\begin{document}

\title{Experimental evidence of above-threshold photoemission in solids}

\author{Francesco Banfi}

\affiliation{Istituto Nazionale per la Fisica della Materia and
Dipartimento di Matematica e Fisica,\\ Universit\`a Cattolica,
I-25121 Brescia}

\affiliation{Istituto Nazionale per la
Fisica della Materia and Dipartimento di Fisica A. Volta,\\
Universit\`a di Pavia, I-27100 Pavia}

\author{Claudio Giannetti}
\author{Gabriele Ferrini}
\author{Gianluca Galimberti}
\author{Stefania Pagliara}
\author{Daniele Fausti}
\author{Fulvio Parmigiani}

\affiliation{Istituto Nazionale per la Fisica della Materia and
Dipartimento di Matematica e Fisica,\\ Universit\`a Cattolica,
I-25121 Brescia}


\preprint{UCSC-Oct03}

\begin{abstract}

Nonlinear photoemission from a silver single-crystal is
investigated by femtosecond laser pulses in a perturbative regime. A clear observation of
above-threshold photoemission in solids is reported for the first time. The ratio between the three-photon above-threshold and the two-photon Fermi edges is found to be 10$^{-4}$. This value constitues the only available test-bench for theories aimed to understand the mechanism responsible for above-threshold photoemission in solids.
\end{abstract}\

\maketitle\

In recent years a significant effort has been dedicated to the study
of above-threshold ionization
(ATI) \cite{Agostini,McIlrath,Freeman,Gillen, Dewitt}.
A photoionization process is defined ATI when the number $n$ of photons of energy $h\nu$ absorbed by an atom to eject one electron is larger than the minimum number $m$ of photons necessary to overcome the ionization potential W$_{ion}$: $mh\nu<nh\nu$ where $(m-1)h\nu<$W$_{ion}$$<mh\nu$.

With the advent of fs laser pulses, the investigation of ATI in atoms evolved from a perturbative \cite{Agostini} toward a non-perturbative regime \cite{Mainfray, Brabec}. In this last case the electromagnetic field, acting on the electrons bounded in the atom, is comparable or greater than the atomic coulomb field. Under this conditions, attosecond high harmonic radiations can be generated \cite{Brabec}.


In spite of the progresses made by ATI experiments in atoms \cite{Agostini,
McIlrath, Freeman, Gillen} and molecules \cite{Dewitt}, its
equivalent in solids, referred as above threshold photoemission
(ATP), lacks of the experimental evidences otherwise present. In solids, the issue has been tackled for a decade without success, the problem being the detection of ATP features using moderate laser pulse intensities ($\sim$10 $\mu$Joule/cm$^{2}$ per pulse), in order to avoid space-charge effects. This conditions set the non-linear photoemission experiments in the perturbative regime. Actually a few studies do claim observations of ATP in solids
\cite{Luan, Fann, Farkas, Farkas2}, but none of them shows spectra where ATP features are clearly
detected and unambiguously identified.

In this Letter we report non-linear photoemission spectra at the Ag(100) surface that unequivocally show a three photon \textit{above-threshold} photoemission of the Fermi edge together with a two-photon Fermi edge. The intensity ratio between the two spectral features is $\sim$10$^{-4}$. This value is a test-bench for theories aimed to clarify the mechanism of above-threshold photoionization in solids. Interestingly, this ratio is two order of magnitude higher than that estimated through a perturbative, dipole-based, calculation. At this light we cannot disregard an explanation in the frame of scattering-mediated photoemission mechanisms \cite{Rethfeld}, otherwise defined as inverse Bremsstrahlung processes.

\begin{figure}[]
\includegraphics[keepaspectratio,clip,width=0.5\textwidth]{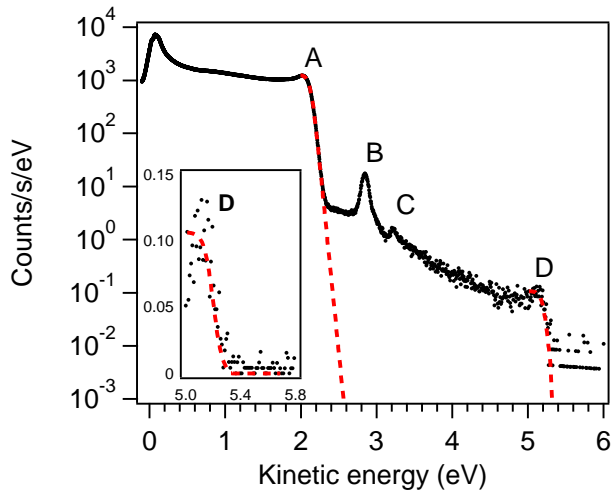}
\caption{Photoemission spectrum from Ag(100) obtained with 150
fs-3.14 eV laser pulses in semilog plot. The light is p-polarized
and its angle of incidence is $30^\circ$ with respect to the
surface normal. The electrons are detected along the surface
normal. The dashed lines (figure and inset) are a Fermi-Dirac fit
at a temperature of 348 K. A temperature slightly in excess of 300
K can be ascribed to the modest space-charge deformation of the
Fermi-edge. Letters A and D label the 2-photon and the 3-photon
Fermi edges respectively. The 2-photon Fermi edge is a commonly observed non-linear emission process, whereas the 3-photon Fermi edge is attributed to an above threshold process. The 3P-AT Fermi edge is reported in
the inset in linear scale. Letters B and C identify the first two
image states.} \label{spectrum}
\end{figure}

Recognition of an above-threshold Fermi-edge should rely on
several independent validations. Indeed, by exciting the sample
with a photon at an energy $h\nu=3.14$~eV, we observe a two-photon
edge and an above threshold three-photon edge resulting from an \textit{above-threshold}
one-photon replica of the two-photon edge. The two- and
three-photon emissions are fitted by the same Fermi-Dirac (FD)
distribution function, see Fig.~\ref{spectrum}. Furthermore, total photoemission yield
measurements show a quadratic and cubic dependence on laser
intensity of the second and third order Fermi edges. This
measurement endorses the photoemission of the above-threshold
Fermi edge as a third order process. A last evidence
resides in an energy shift of the above mentioned features by an
amount of 160 meV and 240 meV respectively, upon a photon energy
change of 80 meV. This measurement can be regarded as a clear
proof that the two Fermi edges are photoemitted by two- and
three-photon processes, the latter resulting from the absorption of an additional photon above-threshold.

The measurements are performed with a laser pulse intensity
$I\sim$ 0.4 GW/cm$^{2}$, at a wavelength of 395 nm. The
angle of incidence is 30$^{\circ}$ with respect to normal incidence. The absorbed average intensity per pulse results, once corrected for
Fresnel losses \cite{nota 2}, in  $\sim$ 60
MW/cm$^{2}$. This implies a Keldysh factor \cite{comment_0} $\gamma\sim $ 1500 , well within the range of $\gamma$-values pertinent to the perturbative regime ($\gamma>$1). Space charge effects are investigated acquiring electron spectra
at different laser intensities, spanning the range 0.1-1
GW/cm$^{2}$ per pulse. A shift of the work function, induced by the space charge, in the non-linear regime is at most
200 meV, when compared to the value obtained from direct photoemission
measurements ($h\nu$=6.28 eV).
Detection of 3P-AT Fermi edge requires extensive statistic. In order to improve the statistic, we accepted a modest space-charge effect that, however, does not preclude fitting of the 3P-AT and 2P Fermi edges with the same Fermi-Dirac distribution, see fits in Fig.\ref{spectrum}. Note worthly, at this
intensity ponderomotive potential is negligible.

The photoemission measurements are performed on a Ag
single-crystal oriented along the (100) surface with an error of
$\pm\,2\,^{\circ}$. The light source is an amplified Ti:Sapphire
laser system emitting linearly polarized pulses with a time width
of 150~fs at 1~kHz repetition rate. The fundamental wavelength can
be tuned in the range 810-790~nm, the second harmonic spans the
photon energy range 3.06-3.14~eV. The experiments are carried out
in a ultrahigh vacuum (UHV) chamber system with a base pressure of
$2\times10^{-10}$~mbar at room temperature. All the measurements
are carried out in normal detection (with respect to the surface
plane) of the photoemitted electrons and excitation at an angle of
incidence of 30$^{\circ}$ in both $p$ and $s$ polarization.
Details of the experimental set-up and sample preparation are
reported elsewhere \cite{Ferrini}.

\begin{figure}[]
\includegraphics[keepaspectratio,clip,angle=-90,width=0.43\textwidth]{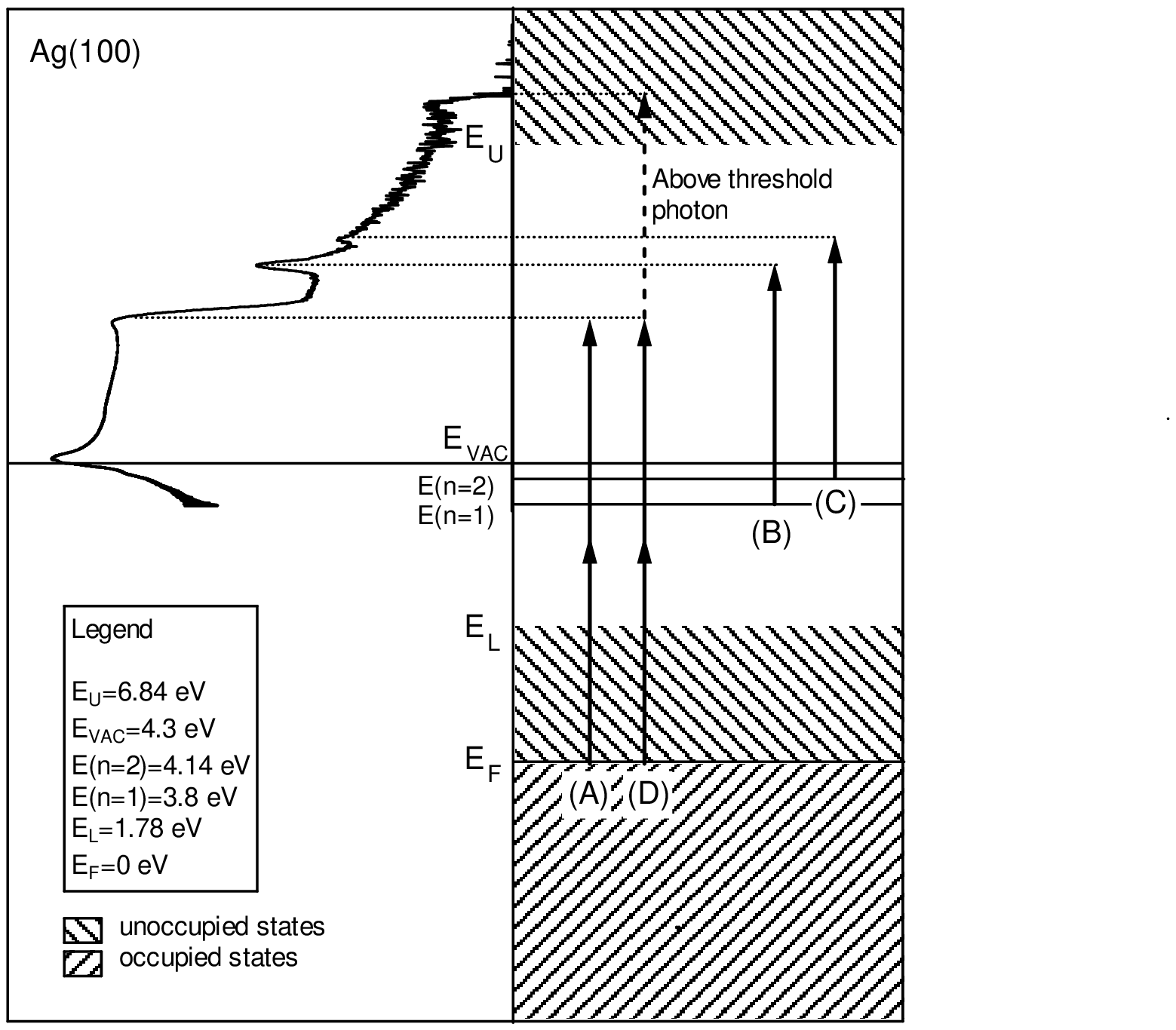}
\caption{Schematic energy diagram for Ag(100) at the $\Gamma$
point and photoemission processes involved in the measured
spectrum. The values reported in the legend are from
Ref.\cite{Fauster}. Labels A, B, C and D are defined as in the
text. The dotted arrow reppresents the \textit{above-threshold} photon absorption} \label{mechanism}
\end{figure}

The spectrum, obtained irradiating with photons of energy
$h\nu$=$3.14$~eV, is shown in Fig.~\ref{spectrum}. Two distinct
portions of the spectrum are clearly distinguishable. The first
one lays in the energy range below 2.3 eV. The second one lays in
the energy range between 2.3 eV and 5.3 eV. We attribute the first
portion of the spectrum to a two-photon photoemission process that
involves electrons occupying the s-p bands lying just below the
Fermi energy, whereas the second portion resembles the low energy
portion upon translation of one photon energy. This evidence is
particularly clear when considering the steps labelled A and D in
the photoemission spectrum. Step A is attributed to the two-photon
Fermi edge, in fact $E_{A}=2h\nu-\Phi$ and the spectrum is well
fitted by a Fermi-Dirac function at 348 K. Step D is energy
shifted by a photon energy $h\nu$ with respect to step A and it
fits a Fermi-Dirac function with the same parameters as the above
mentioned one. The fitting temperature slightly in excess of 300 K
can be ascribed to the modest space-charge induced broadening of
the Fermi edge. We thus identify step D as a three-photon Fermi
edge where a photon in excess of the minimum number strictly
necessary for the photoemission process is absorbed
above-threshold. The peaks labelled B and C are attributed to the
n=1 and n=2 Rydberg-like image potential states (IPS) as a result
of a two-step photoemission process \cite{Ferrini_accepted}. The
measured binding energies are consistent with the expected
theoretical values \cite{Li} and in good agreement with recent
experimental findings \cite{Ferrini}. The present discussion is
schematically illustrated in Fig.~\ref{mechanism}

\begin{figure}[]
\includegraphics[keepaspectratio,clip,width=0.5\textwidth]{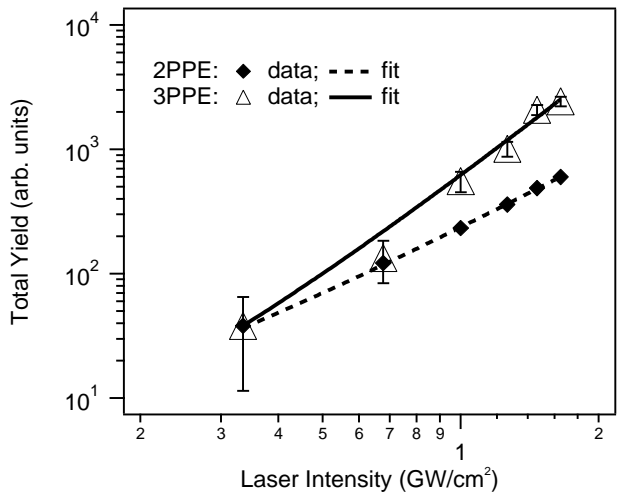}
\caption{Integrated electron yield versus laser fluence at a
photon energy of 3.14~eV. The empty triangles refers to data
obtained integrating the photoemission spectra in the energy range
between 0.41 eV and 2.11~eV; the full triangles  refers to data
obtained integrating the photoemission spectra in the energy range
between 3.55 and 5.25~eV. The latter data are multiplied by a
factor of $2.6\times10^{5}$ for ease of visibility. The full and
dashed line are polynomial fits indicating a second and third
order power dependence on intensity respectively.} \label{yield}
\end{figure}

To further support this interpretation, the integrated
electron yield is measured at different laser intensities. The results are
reported in Fig.~\ref{yield}. The integrated yield in the energy
range between 0.41 and 2.11 eV scales as $I^{2}$, whereas the
integrated yield in the energy range between 3.55 and 5.25~eV
scales as $I^{3}$. The integrations on the Fermi distribution-like
portions of the spectrum are performed on energy ranges
shifted by one photon energy with respect to each other. These
findings are consistent with a second and third-order Fermi edge spectral feature.

\begin{figure}[]
\includegraphics[keepaspectratio,clip,width=0.5\textwidth]{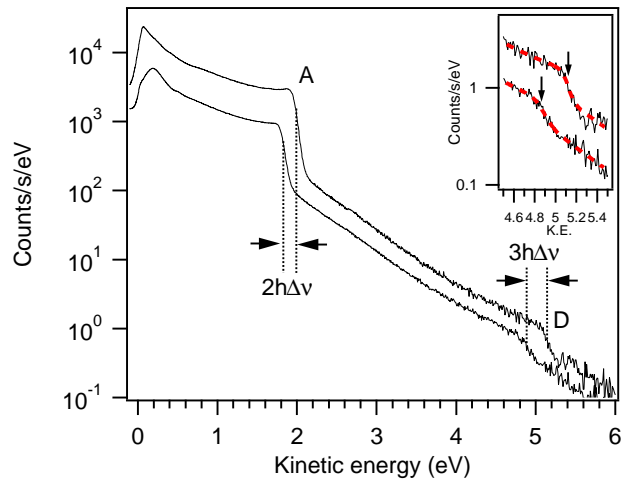}
\caption{Photoemission spectra from Ag(100) obtained with 150 fs
s-polarized laser pulses at $h\nu=3.14$ eV (upper curve) and
$h\nu=3.06$ eV (lower curve) in semilog plot. The angle of
incidence is $30^\circ$ with respect to the surface normal. The
electrons are detected along the surface normal. The impinging
intensity is 0.14 GW/cm$^2$. The plots are displaced vertically
for ease of visualization. In the inset the curves, displaced
vertically, are shown with a fit obtained from a sum of a
Fermi-Dirac and a Maxwell-Boltzmann function convoluted with a
gaussian function. The Maxwell-Boltzmann function fits the "`hot electron tail"' of the spectrum, as explained in Ref. \cite{Banfi}, whereas the experimental apparatus resolution is taken into account via the convolution with a gaussian function, see Ref. \cite{Giannetti} for details.} \label{newVSold}
\end{figure}

A decisive proof is the measurement of the
kinetic energy shift of the two and three-photon Fermi edges upon a
frequency shift $h\Delta\nu$=80 meV of the impinging radiation. Changing the photon energy from
3.14 eV to 3.06 eV, we observe an energy shift of the edges A and
D of the spectrum by an amount of 2$h\Delta\nu$=160 meV and
3$h\Delta\nu$=240 meV respectively, implying that edges A and D
are emitted by a two and three-photon process respectively as shown
in Fig.~\ref{newVSold}

To properly compare ATI in gasses versus solid, we consider ATI in gasses only within the perturbative regime. In this frame, ATI spectra in atoms consist in several peaks of increasing order where the intensity of successive emissions are comparable \cite{McIlrath}. Differently, on the solid, we observe only one spectral feature photoemitted above-threshold with an intensity four order of magnitude lower as compared to the emission of second order. Therefore, on the solid, a direct dipole-transition process cannot be, a-priori, ruled out. In a direct dipole transition the differential cross-section for an n-photon absorption
involving the initial and final states $\left|i\right\rangle$ and
$\left|f\right\rangle$ is, writing the interaction hamiltonian in the \textit{velocity gauge},

\begin{equation}
\frac{dW^{\left(n\right)}_{i\rightarrow f}\left(\mathbf k_{f}\right)}{d\Omega}=\frac{2\pi}{\hbar^{2}}\left(\frac{e^{2}}{2m^{2}c\epsilon_{0}}\frac{I}{\omega^2}\right)^n \left|T^{\left(n\right)}_{i\rightarrow f }\left(\mathbf k_{f}\right)\right| ^{2}\rho\left(E_{k_{f}}\right),
\label{crossect}
\end{equation}

where $\epsilon_{0}$ is the electric permittivity, $e$ the
electron charge, $\rho\left(E_{k_{f}}\right)$ the free electron
final density of states evaluated at the photoemitted electron
energy $E_{k_{f}}=nh\nu-\Phi$, $\mathbf{k_{f}}$ being the wave
vector of the electron in the final state, whose modulus is
determined from energy conservation
$\hbar^{2}k^{2}_{f}/2m=n\hbar\omega-\Phi$. $I$ is the
\textsl{absorbed} laser intensity. The transition matrix element
remains defined as

\begin{equation}
T^{\left(n\right)}_{i\rightarrow f}\equiv\left\langle f\left|p\underbrace{G\left(E_{i}+\left(n-1\right)\hbar\omega \right)p}_{n-1\; such\; terms}...G\left(E_{i}+\hbar\omega \right)p\right|i\right\rangle
\label{tmatrix}
\end{equation}

with $p=\mathbf{\epsilon}\cdot\mathbf{p}$, $\mathbf\epsilon$ being
the polarization unit vector of the e.m field $\mathbf{E}$,
$\mathbf{p}$ the momentum operator and
$G\left(E_{i}+m\hbar\omega\right)$ the propagator of the
unperturbed hamiltonian after absorption of the $m^{th}$ photon. The eigenvalues $E_{i}$ and the corresponding eigenvectors, entering the perturbative calculation, are the solutions of the unperturbed semi-infinite crystal hamiltonian. We pinpoint that the above threshold transition involves the mixing of the final free electron state with
\textit{all} the unperturbed hamiltonian eigenstates and not only with the
eigenstates entering the continuum spectrum, that would give null
matrix elements. An estimate
for the ratio $\partial_{\Omega}W^{\left(3\right)}_{i\rightarrow f}\left(\mathbf
k_{f}\right)/\partial_{\Omega}W^{\left(2\right)}_{i\rightarrow
f}\left(\mathbf k_{f}\right)$, among the 3-photon and the 2-photon
differential cross-sections, is $\sim 10^{-6}$. The calculation was carried out neglecting the
details of the matrix elements, within the same approximations
used in \cite{Faisal}:

\def\sumint{\mathop{\sum\hspace{-0.5cm}\int}}

\begin{equation}
\frac{T^{\left(3\right)}_{i\rightarrow f }\left(\mathbf k_{f}\right)}
{T^{\left(2\right)}_{i\rightarrow f }\left(\mathbf k_{f}\right)}\approx
\sumint_{j}\frac{p}{\left(E_{i}-E_{j}\right)+2\hbar\omega},
\label{estimate}
\end{equation}

$p \sim \hbar k_{f}=\sqrt{2m(3\hbar\omega-\Phi)}$. Assuming that
only one particular eigenvalue ${E}_{j}=\tilde{E}$ is relevant in
the sum, the right term in Eq.~\ref{estimate} reduces to
$p/\left(\Delta E + 2\hbar\omega\right)$, where $\Delta E \equiv
E_{i}-\tilde{E}$. We take $\Delta E$ $\sim$ -2 eV, meaning that the prevailing
unperturbed eigenstates contributing to the transition matrix are
the unoccupied s-p bulk states laying $\sim$ 8 eV above the Fermi level. The occurence of resonances is unlikely, our statement being based on the reported band-structure for Ag at the $\Gamma$ point. In spite of the approximations used for the present model, as well as for other models reported in the litterature \cite{Georges}, the difference between the calculated ratio ($\partial_{\Omega}W^{\left(3\right)}_{i\rightarrow f}\left(\mathbf
k_{f}\right)/\partial_{\Omega}W^{\left(2\right)}_{i\rightarrow
f}\left(\mathbf k_{f}\right)$ $\sim$ 10$^{-6}$), when compared with the experimental ratio (3P-AT Fermi edge/2P Fermi edge $\sim$ 10$^{-4}$) could be significant as for the physics involved.

An alternative explanation takes into account the possibility of scattering mediated light absorption. Such processes have been recently proposed by Lugovskoy et al. \cite{Lugovskoy} and Rethfeld et al. \cite{Rethfeld} to account for the laser induced intraband absorption in metallic s-p conduction bands. The scattering mechanism is attributed to electron-phonon or electron-ion collisions (inverse Bremsstrahlung). The inverse Bremsstrahlung process accounts for momentum conservation in an s-p intraband electron transition and it constitutes the main absorption channel at laser intensities of $\sim$ 10$^{2}$ MW/cm$^{2}$. Solution for the non-equilibrium electron distribution for a metal under laser irradiation, where the above mentioned scattering mechanisms are dominant, results in a step-like distribution with steps separated by a single photon energy. The electronic distribution strongly resembles the photoemission spectra reported in this Letter (see Fig.~1b in \cite{Rethfeld}). Unfortunatly, only qualitative comparisions are possible between our data and the inverse Bremsstrahlung-based model. On the other hand, the present experimental data are, to the best of our knowledge, the only reliable to be compared with theories.

In conclusion we report unambiguous evidence of
an \textit{above threshold} photoemitted Fermi edge in solids. The experiment is carried out impinging with 10 $\mu$Joule/cm$^{2}$ per pulse, to avoid space-charge effects, setting the photoemission process is in the perturbative regime, $\gamma\gg1$. The three-photon above-threshold Fermi edge is identified on the base of three independent experiments. The experimental ratio between the 3P-AT Fermi edge and the 2PPE is $\sim$ 10$^{-4}$. This ratio seems to be significantly higher than expected according to non-linear perturbative direct dipole transition, while it agrees only qualitatively with recently reported inverse Bremsstrahlung models. At this light it is clear than any further improvement of the actual ATP theory must be compared with the measurements reported in this Letter.

\end{document}